\documentclass{webofc}
\usepackage[varg]{txfonts}   
\pdfoutput=1

\usepackage{epsfig}
\usepackage{graphicx}
\usepackage{xcolor}
\usepackage{caption}
\usepackage{float}
\usepackage{framed}

\usepackage{latexsym,times}
\usepackage{amssymb,amsmath}
\usepackage{amsxtra}
\usepackage{amscd}
\usepackage{amsfonts}

\usepackage{hyperref}

\newcommand{\half}{{{\textstyle\frac{1}{2}}}}
\newcommand{\quarter}{{{\textstyle\frac{1}{4}}}}
\newcommand{\be}{\begin{equation}}
\newcommand{\ee}{\end{equation} }
\newcommand{\beqa}{\begin{eqnarray} }
\newcommand{\eeqa}{\end{eqnarray} }
\newcommand{\ba}{\begin{array}}
\newcommand{\ea}{\end{array}}
\newcommand{\bpm}{\begin{pmatrix}}
\newcommand{\epm}{\end{pmatrix}}

\newcommand{\Spin}{\mathbf{Spin}}

\newcommand{\ODD}{\mathbf{O}(D,D)}

\newcommand{\SpinD}{{\Spin(1,D{-1})}}
\newcommand{\oSpinD}{{{\Spin}(D{-1},1)}}

\newcommand\rd{{\rm d}}
\newcommand\rD{{\rm D}}

\newcommand\cA{{\cal A}}

\newcommand\cD{{\cal D}}

\newcommand\cH{{\cal H}}

\newcommand\cJ{{\cal J}}

\newcommand\dis{\displaystyle}

\def\tx{\tilde{x}}

\def\tpartial{\tilde{\partial}}

\def\breta{\bar{\eta}}

\def\brpsi{\bar{\psi}}

\def\brp{{\bar{p}}}
\def\brq{{\bar{q}}}

\def\brPhi{{{\bar{\Phi}}}}

\def\brV{{\bar{V}}}

\def\brP{{\bar{P}}}

\newcommand{\na}{{\nabla}}

\newcommand{\red}[1]{{\color{red} #1 \color{black}}}

\newcommand{\darkgreen}[1]{\textcolor[rgb]{0.1,0.3,0.1}{#1}}

\begin{document}
\title{\vspace{-30pt}\mbox{Stringy Gravity:  Solving the Dark Problems at `short' distance}}
%
%

\author{\firstname{Jeong-Hyuck} \lastname{Park}
\thanks{\email{park@sogang.ac.kr}} }

\institute{Department of Physics, Sogang University, 35 Baekbeom-ro, Mapo-gu,  Seoul  04107, Korea 
\and
         Center for Theoretical Physics of the Universe, Institute for Basic Science (IBS), Seoul 08826,  Korea}

\abstract{Dictated by Symmetry Principle,  string theory predicts  \textit{not} General Relativity \textit{but} its own gravity which assumes  the entire closed string   massless sector     to be  geometric and thus gravitational.  
In terms of  $R/(MG)$, \textit{i.e.~}the dimensionless radial variable  normalized by mass,    Stringy Gravity agrees with General Relativity  toward    infinity, but modifies it at short  distance. At far short   distance, gravitational force can be  even  repulsive. These may solve the dark matter and energy problems, as they  arise   essentially     from  small $R/(MG)$ observations: long distance divided by much heavier mass.  We address        the pertinent 
differential geometry for Stringy Gravity,  stringy Equivalence Principle,  stringy geodesics and the minimal coupling to the Standard Model.     We highlight the notion of   `doubled-yet-gauged'   coordinate system,  in which a gauge orbit corresponds to   a single  physical  point and proper distance is  defined between two gauge orbits by a path integral.  
}
\maketitle

\begin{center}
\vspace{9pt}
\begin{figure}[H]
\centering
\includegraphics[width=62mm]{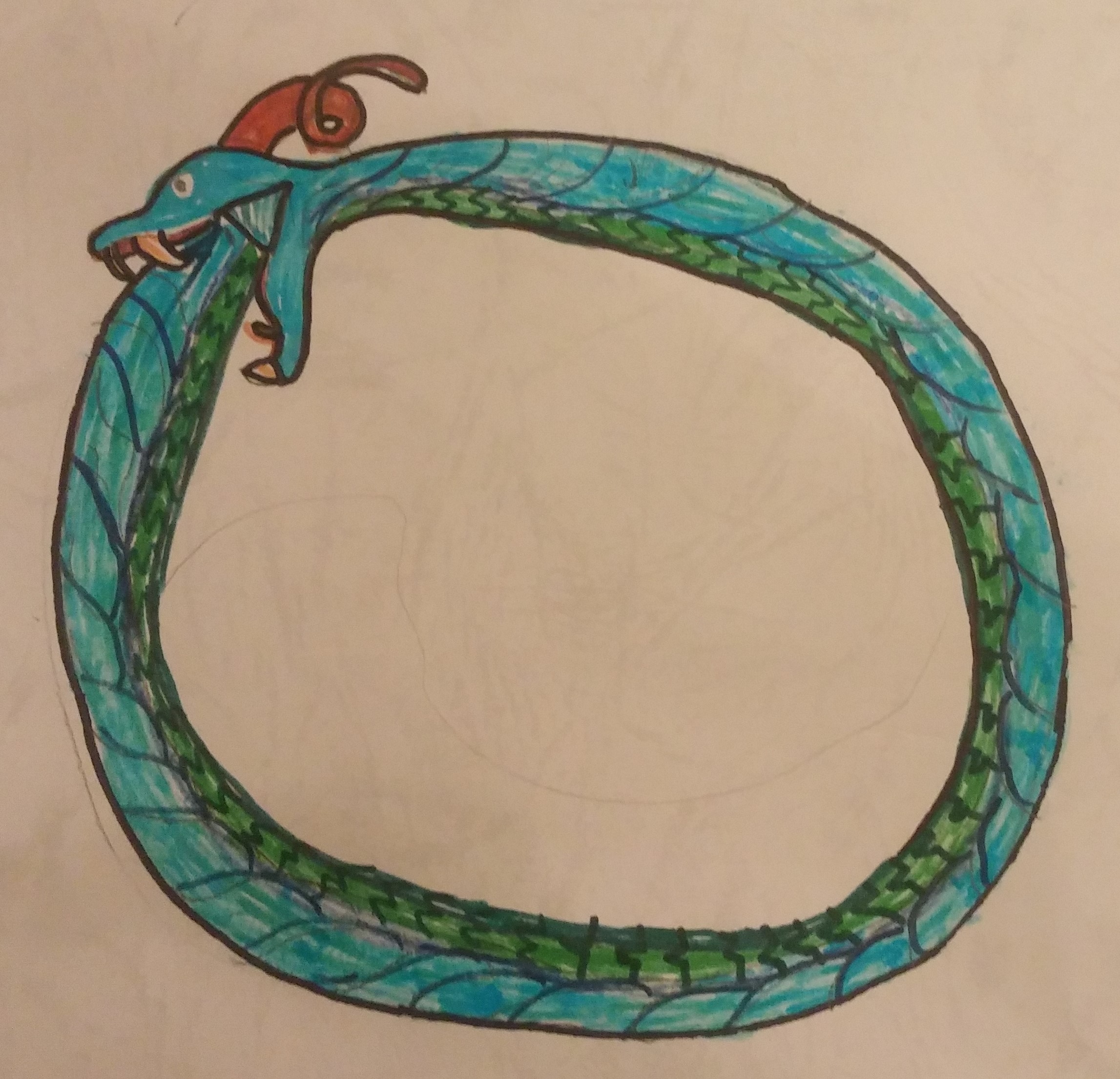}
\caption*{\darkgreen{\bf{\it{{Uroboros\,: An ancient Egyptian symbol for a serpent which eats its own tail }}}}}
\captionsetup{labelformat=empty}
\end{figure}
\vspace{9pt}

\mbox{\sc\red{\href{https://www.apctp.org/plan.php/ICGAC-IK}{Proceeding of   International Joint Conference  of\, ICGAC-XIII and IK15}}}\\
\red{Also based on talks  in \href{http://www.birs.ca/events/2017/5-day-workshops/17w5018}{Banff}, \href{http://home.kias.re.kr/MKG/h/IbsKias2017/?pageNo=2683}{High\,1}, \href{http://home.kias.re.kr/MKG/h/WFSG2017/?pageNo=2761}{Seoul\,1}, \href{http://webs.um.es/jj.fernandezmelgarejo/geom17/}{Kyoto} , \href{http://thphys.irb.hr/dualities2017/?home}{Zagreb}, \href{https://www.kasi.re.kr/eng/post/eng_colloquium/9986}{KASI}, \href{https://sites.google.com/site/khuworkshop/home}{Seoul\,2},  \href{http://gw.phy.cuhk.edu.hk/cod/}{Hong Kong}.} 
\end{center}

\newpage

\section{Introduction}
\label{intro}Ever since  Einstein formulated his theory of gravity, \textit{i.e.~}General Relativity (GR),     the  metric, $g_{\mu\nu}$, has been privileged to be   the only  {geometric}   and hence  {gravitational} field, on account of   the adopted    Riemannian  geometry.  All other fields or particles    are meant to  be     `extra' matters. In particular,  the Riemannian metric sets a  proper length which  every   free  particle would minimize, following the geodesic,
\be
\ddot{x}^{\lambda}+\left\{{}^{~\lambda}_{\mu\,\nu}\right\}
\dot{x}^{\mu}\dot{x}^{\nu}=0\,.
\label{geodesic}
\ee

The   gravitational coupling  to  matters, or to the Standard Model,   is  `minimally' fixed  by the   metric and   covariant derivatives   which ensure both  diffeomorphisms and  local Lorentz symmetry. In the words of Cheng-Ning Yang, {\textit{\red{symmetry dictates interaction.}}}

On the other hand,   string theory  suggests  us to put a  two-form gauge potential, $B_{\mu\nu}$, and a scalar dilaton, $\phi$, on an equal footing along with the metric. Forming the  massless sector of closed strings, they are ubiquitous in all  string theories, with the  conventional  (Riemannian) description given by  
\be
\ba{lll}
\dis{\int\rd^{D}x}~\sqrt{-g}e^{-2\phi}\left(R_{g}+4\partial_{\mu}\phi\partial^{\mu}\phi
-\textstyle{\frac{1}{12}}H_{\lambda\mu\nu}H^{\lambda\mu\nu}\right)\quad&~\mbox{where}~&\quad H=\rd B\,.
\ea
\label{NSNS}
\ee
Furthermore, a genuine stringy  symmetry, called T-duality, can  mix three of them~\cite{Buscher:1987sk,Buscher:1987qj}. These  may well hint at the existence of    \textit{{Stringy Gravity}} which should take  the entire   closed string massless sector  to be  geometric  and  gravitational. For this,  a   novel differential geometry beyond Riemann is required, in particular to  shed  light on the 
  $\ODD$ T-duality symmetry hidden in  the   above  action.

 After series of pioneering works on `doubled sigma models'~\cite{Duff:1989tf,Tseytlin:1990nb,Tseytlin:1990va,Hull:2004in,Hull:2006qs,Hull:2006va} and `double field theory' (DFT)~\cite{Siegel:1993xq,Siegel:1993th,Hull:2009mi,Hull:2009zb,Hohm:2010pp} (\textit{c.f.~}\cite{Aldazabal:2013sca,Berman:2013eva,Hohm:2013bwa}),  such an idea of Stringy Gravity has    materialized.  The word  `double' refers to the fact that  doubled $(D\,{+D})$-dimensional coordinates are  used for the description of  $D$-dimensional physical  spacetime. While   such a usage was  historically first made   in the case of a torus background  --    by introducing a dual coordinate conjugate to the  string   winding momentum --  the  doubled coordinates turned out to be far   more general: they  can be applied to any compact or  non-compact spacetime, as well as to   not only string but also  to  particle physics.


\section{Stringy Gravity: Uroboros solution to Dark Problems}
\label{sec1}
\subsection{Stringy differential geometry}
\label{subsec1}

Stringy Gravity of our interest  adopts    the \textit{doubled-yet-gauged coordinate system}~\cite{Park:2013mpa} which meets   two  properties.  Firstly,   an  $\ODD$ group is  \textit{a priori}   postulated,   having the constant  invariant ``metric",   
\be
\cJ_{AB}={\small{\left({\mathbf{\ba{cc}\bf{0}&\bf{1}\\\bf{1}&\bf{0}\ea}}\right)}}\,.
\label{cJ}
\ee
 With the inverse,  $\cJ^{AB}$, 
it can freely raise and lower the $\ODD$ vector indices (capital letters, $A,B,\cdots$). Secondly,     the doubled coordinates are   gauged by  an   equivalence relation,
\be
x^{A}~\sim~x^{A}+\Delta^{A}(x)\,,
\label{ER}
\ee
where $\Delta^{A}$ is an arbitrary `derivative-index-valued'  vector. This   means that its superscript index must be identifiable  as that of   derivative, $\partial^{A}=\cJ^{AB}\partial_{B}$. For example, with arbitrary functions, ${\Phi}_{i}$,  belonging to the theory,  $\Delta^{A}=\sum_{i,j}\Phi_{i}\partial^{A}\Phi_{j}$.  
The equivalence relation~(\ref{ER}) can be    realized by  requiring that every  field or function in Stringy Gravity -- such as $\Phi_{i}$ itself,  physical fields,  local symmetry parameters   and  their arbitrary  derivatives --  should be  invariant under the coordinate gauge symmetry shift,\footnote{See also Eq.\eqref{gaugedDx} for the cases  where the doubled coordinates are dynamical fields.}
\be
\Phi_{k}(x+\Delta)=\Phi_{k}(x)\quad\Longleftrightarrow\quad\Delta^{A}\partial_{A}=0\,.
\label{CGS}
\ee 
In this way,  it becomes   not a point but   a gauge orbit  in the doubled coordinate system   that  corresponds to  a single physical point. The above coordinate gauge symmetry  invariance is equivalent to so called  the `section condition' in DFT, 
\be
\partial_{A}\partial^{A}\,{=0}\,.
\ee 
With respect to the off block-diagonal form of the $\ODD$  metric~(\ref{cJ}), the doubled coordinates split into two parts: $x^{A}=(\tilde{x}_{\mu}, x^{\nu})$, and hence $\partial_{A}\partial^{A}=2\tpartial^{\mu}\partial_{\mu}$. 
The general solution to the section condition is then   given by $\tpartial^{\mu}\,{\equiv0}$, up to $\ODD$ rotations~\cite{Siegel:1993th,Hull:2009mi}.

General covariance in doubled-yet-gauged spacetime, or doubled diffeomorphisms,  are given by
\be
\ba{cc}
\delta x^{A}=V^{A}\,,
\quad&\quad
\delta\partial_{A}=-\partial_{A}V^{B}\partial_{B}=(\partial^{B}V_{A}-\partial_{A}V^{B})\partial_{B}\,,
\ea
\label{diff1}
\ee
and  for a covariant tensor (or tensor density with weight $\omega$),\footnote{Eq.\eqref{diff2} corresponds to the passive counterpart  of  ``generalized Lie derivative" \textit{\`{a} la} Siegel~\cite{Siegel:1993th}. }
\be
\delta T_{A_{1}\cdots A_{n}}=-\omega\partial_{B}V^{B}T_{A_{1}\cdots A_{n}}+\textstyle{\sum_{i=1}^{n}\,}(\partial_{B}V_{A_{i}}-\partial_{A_{i}}V_{B})T_{A_{1}\cdots A_{i-1}}{}^{B}{}_{A_{i+1}\cdots  A_{n}}\,.
\label{diff2}
\ee

 The whole  massless sector of closed strings, or   \textit{{stringy gravitons}},  consist of 
 a unit-weighted  scalar density given by an exponentiation of   DFT-dilaton,  $e^{-2d}$,  and a symmetric projector,  
\be 
\ba{ll}
P_{AB}=P_{BA}\,,\quad&\quad
P_{A}{}^{B}P_{B}{}^{C}=P_{A}{}^{C}\,,
\ea
\ee
which accompanies   the  complementary,  orthogonal projector, $\brP_{AB}=\cJ_{AB}-P_{AB}$. It follows that     the difference of the two projectors   sets a symmetric $\ODD$ element, \textit{i.e.~}DFT-metric,  $P_{AB}-\brP_{AB}=\cH_{AB}$, satisfying $\cH_{AB}=\cH_{BA}$ and $\cH_{A}{}^{C}\cH_{B}{}^{D}\cJ_{CD}=\cJ_{AB}$. 

 The above $\ODD$ covariant  defining properties of the  stringy gravitons  can be   conventionally  solved   by $\{g_{\mu\nu}, B_{\mu\nu}, \phi\}$. However, not all the solutions can be parametrized by these GR variables.  Stringy Gravity  is more general than GR: it encompasses   `non-Riemannian'  geometries   where  the Riemannian  metric, $g_{\mu\nu}$,   cannot be   defined,  even locally (see \cite{Lee:2013hma,Ko:2015rha,Park:2016sbw} for examples and \cite{Morand:2017fnv} for the complete classification and relations to Newton-Cartan/Carroll  gravities, chiral strings \textit{etc.}).

Requiring  compatibility  with  all the stringy gravitons, \textit{i.e.~}$\na_{A}d=0$, $\na_{A}P_{BC}=0$, $\na_{A}\brP_{BC}=0$,   and  imposing  some  torsionless conditions, one can   uniquely determine  the  \textit{connection} for  a   diffeomorphism covariant derivative, $\na_{A}=\partial_{A}+\Gamma_{A}$~\cite{Jeon:2011cn},
\be
\textstyle{
\ba{ll}
\Gamma_{CAB}=&2\left(P\partial_{C}P\brP\right)_{[AB]}
+2\left({{\brP}_{[A}{}^{D}{\brP}_{B]}{}^{E}}-{P_{[A}{}^{D}P_{B]}{}^{E}}\right)\partial_{D}P_{EC}\\
{}&-4\left(\textstyle{\frac{1}{P_{M}{}^{M}-1}}P_{C[A}P_{B]}{}^{D}+
\textstyle{\frac{1}{\brP_{M}{}^{M}-1}}\brP_{C[A}\brP_{B]}{}^{D}\right)\!\left(\partial_{D}d+(P\partial^{E}P\brP)_{[ED]}\right)\,.
\ea}
\label{Gammao}
\ee
This stringy analogue of the Christoffel symbol    constitutes subsequently a scalar curvature, $S_{\scriptscriptstyle{(0)\,}}$,  and a two-indexed  ``Ricci'' curvature,  $P_{A}{}^{C}\brP_{B}{}^{D}S_{CD}$,   which further  leads to  the  
stringy counterpart of the  conserved   Einstein   tensor:  $~G_{AB}=4P_{[A}{}^{C}\brP_{B]}{}^{D}S_{CD}-\half\cJ_{AB}
S_{\scriptscriptstyle{(0)}}$ satisfying  $\na_{A}G^{AB}=0$~\cite{Park:2015bza}. 

The scalar curvature multiplied by  the unit-weighted integral measure, $e^{-2d}S_{\scriptscriptstyle{(0)\,}}$, sets then the action for   \textit{pure} Stringy Gravity,   which agrees with  (\ref{NSNS}) upon   Riemannian parametrization and up to total derivatives.  Its  equations of motion  amount to  the vanishing of the  stringy Einstein tensor, $G_{AB}\,{=0}$, or equivalently $S_{\scriptscriptstyle{(0)}}=0$ and $P_{A}{}^{C}\brP_{B}{}^{D}S_{CD}=0$.  In particular,  constant     background solutions   include not only the   flat Minkowskian  spacetime but also non-Riemannian ones, \textit{e.g.}  $\cH_{AB}=\cJ_{AB}$, 
\textit{etc.}~\cite{Morand:2017fnv}.

\subsection{Minimal coupling to the Standard Model}
\label{subsec2}
To couple to  fermions, it is necessary to introduce    \textit{a pair of vielbeins},  $V_{Ap}$, $\brV_{A\brp}$,  which  form  the mutually  orthogonal  projectors,  $P_{AB\,}{=V_{Ap}V_{B}{}^{p}}$,  $\brP_{AB\,}{=\brV_{A\brp}\brV_{B}{}^{\brp}}$, and  satisfy their own defining properties:
\be
\ba{llll}
V_{Ap}V^{A}{}_{q}=\eta_{pq}\,,\quad&\quad
\brV_{A\brp}\brV^{A}{}_{\brq}=\breta_{\brp\brq}\,,
\quad&\quad 
V_{Ap}\brV^{A\brp}=0\,,\quad&\quad
V_{Ap}V_{B}{}^{p}+\brV_{A\brp}\brV_{B}{}^{\brp}=\cJ_{AB}\,.
\ea
\ee 
The unbarred and barred lowercase  letters are    the   vector indices of twofold local Lorentz symmetries,  $\SpinD$ and $\oSpinD$,  subject to  Minkowskian  metrics,  $\eta_{pq}$ (mostly plus) and  $\breta_{\brp\brq}$ (mostly minus),  respectively.  The vielbeins    diagonalize $\cJ_{AB}$ and $\cH_{AB}$ simultaneously, with the eigenvalues, $(\eta,+\breta)$ and $(\eta,-\breta)$.  Physically, the doubling of the spin groups implies   a pair of locally inertial frames   which exist separately  for the    left and  right closed string modes~\cite{Duff:1986ne}.\footnote{Ensuring the twofold  spin  groups in the maximally supersymmetric  DFT~\cite{Jeon:2012hp} and the doubled-yet-gauged  Green-Schwarz superstring action~\cite{Park:2016sbw}, the conventional IIA and IIB theories are unified into a single  unique  theory that is chiral with respect to both spin groups.  The distinction of  IIA and IIB then refers to   two different types of  (Riemannian)  `solutions' rather than `theories'.}   Now,    to take care of all the local symmetries in  Stringy Gravity,  the covariant derivative  generalizes to $\na_{A}=\partial_{A}+\Gamma_{A}+\Phi_{A}+\brPhi_{A}$, where the last two are $\SpinD$ and $\oSpinD$  connections. These  are   fixed  by  the compatibility   with the   vielbeins,   $\na_{A}V_{Bp}\,{=0}$, $\na_{A}\brV_{B\brp}\,{=0}$, and thus related to  the stringy Christoffel symbol~(\ref{Gammao})~\cite{Jeon:2011vx}.

With the basics of the stringy differential geometry, $\{d, V_{Ap}, \brV_{A\brp}, \na_{A}, S_{\scriptscriptstyle{(0)}}\}$,     the Standard Model can be straightforwardly,  and minimally,  coupled to  Stringy Gravity~\cite{Choi:2015bga}.  In particular,   fermionic kinetic terms read  
\be
\!e^{-2d}\,\brpsi\gamma^{A}\na_{A}\psi=e^{-2d}\,\brpsi V^{A}{}_{p}\gamma^{p}({\partial}_{A}\psi+\quarter{\Phi}_{Apq}\gamma^{pq}\psi)\equiv\sqrt{-g}\,\bar{\chi}\gamma^{\mu} \big( \partial_{\mu} \chi + \textstyle{\frac{1}{4}} \omega_{\mu  pq} \gamma^{pq} \chi +\textstyle{\frac{1}{24}} H_{\mu pq} \gamma^{pq} \chi \big)\,.
\label{minimalcouplingF}
\ee
The second equality, or `$\,\equiv\,$',   holds up to the  Riemannian reduction of $\tpartial^{\mu}\,{\equiv0}$,  and a  field redefinition of the  fermion,  $\chi\equiv 2^{-\frac{1}{4}}e^{-\phi}\psi$, which couples to the Riemannian  vierbein and $B$-field, as if fundamental string. That is to say,  dilaton $\phi$  is \textit{dark}  to the fermion, $\chi$.\footnote{On the other hand, gauge bosons couple to $g_{\mu\nu}$ and $\phi\,$  only: $H$-flux is \textit{dark} to the  gauge bosons.}  Although quantum field theory is supposed to describe  ``point-like" particles,  the Dirac spinor $\chi$ seems to remember  its stringy origin.

 From Stringy Gravity point of view, $H$-flux is part of the torsionless connection,  and  the stringy Christoffel symbol~\eqref{Gammao} cannot be  transformed  point-wise to vanish    by  doubled diffeomorphisms. That is to say,  locally inertial (normal) frames  do  not exist in Stringy Gravity:  the Equivalence Principle is  broken for  a string. This should not be a surprise, since string is an extended object, subject to ``tidal force'',  or $H$-flux.  It is also consistent with the absence of a  four-indexed ``Riemann'' curvature in Stringy Gravity~\cite{Jeon:2011cn,Hohm:2011si}.   Eq.\eqref{minimalcouplingF} then implies   the  violation of the Equivalence Principle for fermions (\textit{c.f~}\cite{Lee:1955vk} or ``fifth force'').    We continue to discuss  the recovery of the principle in section~\ref{subsec4}.

Given  the twofold local Lorentz symmetries, one needs to   decide   the spin group  for each fermion in the Standard Model  after preparing   two sets of gamma matrices.   An open question, which could  be answered experimentally, is whether   the spin groups of     quarks and    leptons are different (or not)~\cite{Choi:2015bga}.

\subsection{Uroboros solution to the Dark Matter and the Dark  Energy problems}
\label{subsec3}
Motivated by the coupling of the Standard Model to the   closed string massless sector,  such as \eqref{minimalcouplingF},  the  most general, asymptotically flat,  spherically symmetric \textit{vacuum}  solution to $D\,{=4}\,$ Stringy Gravity  was studied in \cite{Ko:2016dxa} (see also \cite{Burgess:1994kq}). The solution   lets  the stringy  Einstein   tensor  vanish identically,    $G_{AB}\,{=0}$,  and therefore    corresponds  to the stringy analogue of the Schwarzschild solution to GR.   Yet, it  possesses  three free parameters:   not only  mass, $M$ (measurable at infinity), but also,  say  $H$-flux and dilaton  charge. All of them  ought to    reflect  the intrinsic properties of the implicitly assumed  matter     localized at the center, which deserve,  combined with \eqref{minimalcouplingF}, further explorations, possibly in connection with   `energy condition' in GR.

The generic spherical   solution would   be regarded to  feature     a naked singularity,  if it were interpreted from  GR point of view, \textit{c.f.}~\eqref{NSNS}.  But, strictly   within the framework of  Stringy Gravity, it may be   non-singular: all the stringy  curvature tensors are trivial.  Inevitably, the notion of singularity    depends on  the   differential geometry in use. Investigation into  the  geodesic completeness  is of interest.\footnote{\textit{c.f.~}both (\ref{geodesic}) and  (\ref{Sgeodesic}), as well as cosmic censorship \textit{\`{a} la} Penrose. }

Analysis of the circular geodesic  around the center  in `string  frame' -- to be justified in section~\ref{subsec4} below --  shows  that the  rotation curve generically features a maximum and thus non-Keplerian over a finite range, while becoming asymptotically Keplerian toward  infinity~\cite{Ko:2016dxa}.  Specifically,     with  appropriately  tuned  $H$-flux and dilaton  charge,  in terms of $R/(MG)$, \textit{i.e.~}the dimensionless radial variable normalized by mass,  Stringy Gravity agrees with GR  becoming   Keplerian for large   $R/(MG)\,{\gtrsim  10^{8}}$,  but modifies it at shorter distance, $R/(MG)\,{\lesssim  2\times 10^{6}}$ (\textit{c.f.~}MOND~\cite{Milgrom:1983ca}). At far shorter distance,   gravitational force  can be  even  repulsive. These may solve the dark matter and  energy problems which  are essentially based on  small $R/(MG)$ observations: long distance divided by  much heavier mass.

Namely, \textit{the self-interaction of the  closed string massless sector makes  Stringy Gravity modify  GR at `short' distance  in terms of $R/(MG)$,} such that it may    explain   the observed  galaxy rotation curves, hence  solve  the dark matter problem.  Further, \textit{the repulsive force at far shorter  distance   can  be  responsible for the accelerating  expansion of the Universe,} and thus  solve  the  dark energy problem.

\begin{framed}
\begin{figure}[H]
\centering
\begin{minipage}{.48\textwidth}
\vspace{17pt}
\includegraphics[width=59mm]{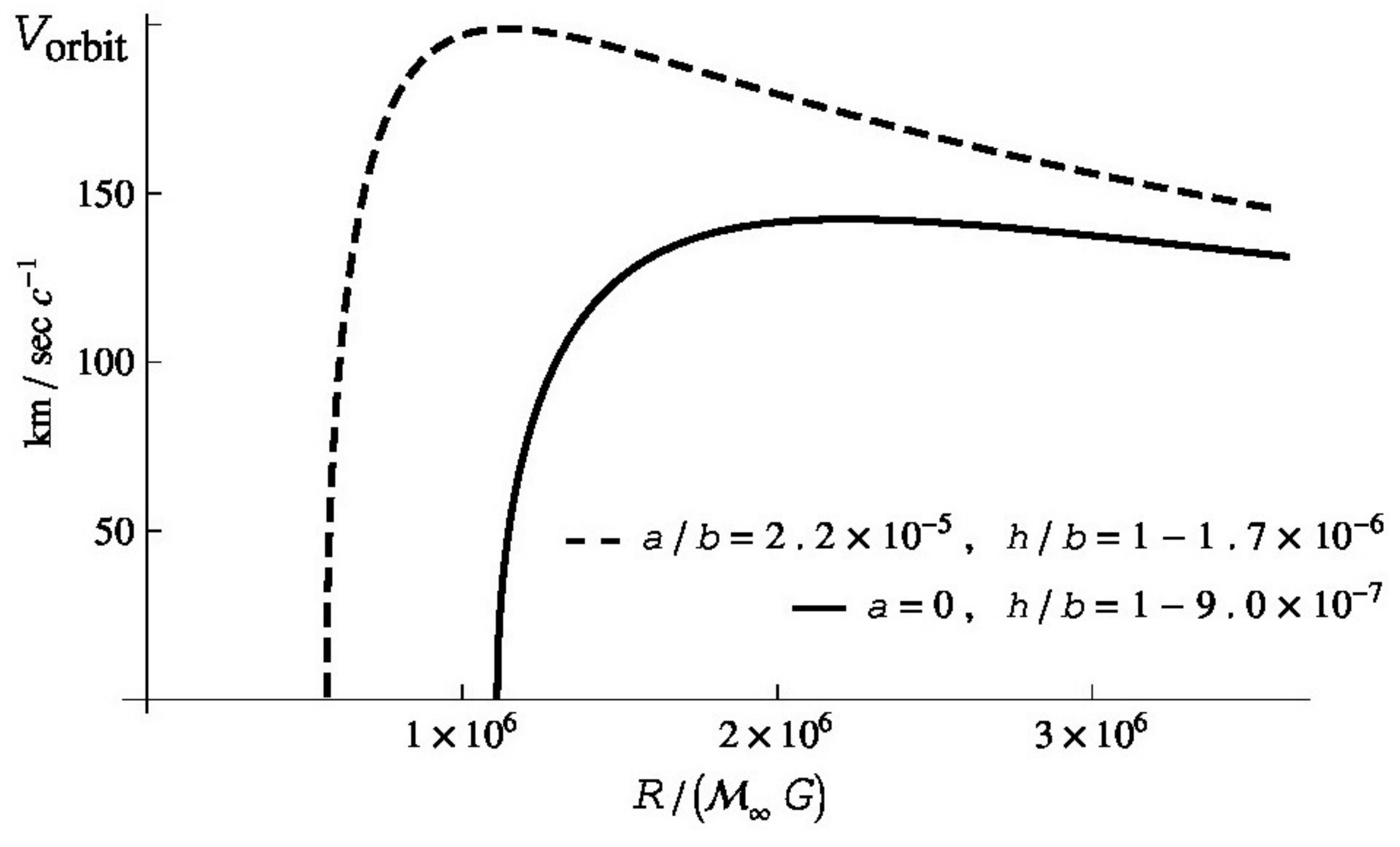}
\captionsetup{labelformat=empty}
\caption{{{\red{Rotation curves in Stringy Gravity~\cite{Ko:2016dxa}}}}}
\end{minipage}%
\begin{minipage}{.52315\textwidth}
  Dimensionless rotation curves, $V_{\rm{orbit}}$ \textit{vs.~}\,$R/(MG)$,  around the spherical vacuum   in Stringy Gravity  feature generically a maximum. By   tuning  the free parameters,  one can make the maximum velocity comparable to \href{https://www.google.co.kr/search?q=galaxy+rotation+curve&newwindow=1&client=firefox-b&source=lnms&tbm=isch&sa=X&ved=0ahUKEwjc7tejvMfNAhXHp5QKHWGUAFIQ_AUICCgB&biw=1280&bih=604}{the observations of galaxy rotations.} Intriguingly as unintended consequence, the horizontal scale, $R/(MG)$,  also  appears    consistent with the observatory data of visible matters, \textit{c.f.~}`Uroboros' spectrum below. For sufficient small $R/(MG)$, the gravitational force can be  repulsive and the orbital velocity is not defined.   Changing  the horizontal axis from $R/(MG)$ to the  physical radius, $R$, the rotation curves can oscillate for nontrivial    radial  mass  distribution functions, $M(R)$.   
\end{minipage}%
\end{figure}
\vspace{-12pt}
\end{framed}
{\begin{center}
\begin{figure}[H]
\centering
\includegraphics[width=142.6mm]{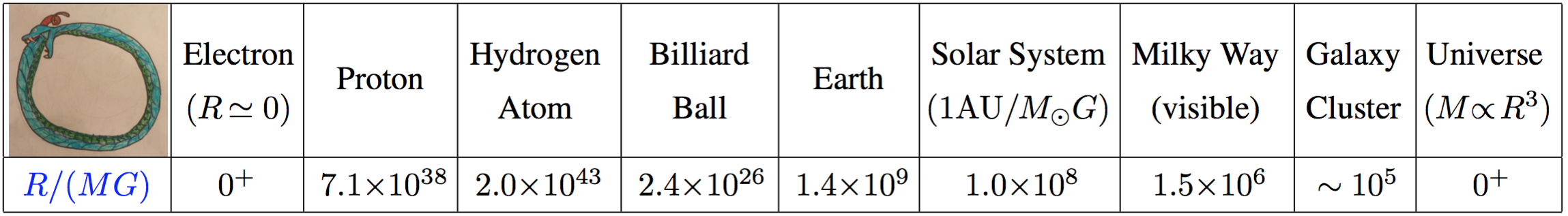}
\caption*{\small{\textbf{\,\qquad `Uroboros' spectrum  of the dimensionless radial variable normalized by mass in natural units. \\
\mbox{\,} Observing stars  and galaxies far  away, we can  probe deep  into the short distance natrue of GRAVITY. }}}
\captionsetup{labelformat=empty}
\end{figure}
\end{center}
}

\subsection{Proper distance in doubled-yet-gauged spacetime: Justification for  String Frame}
\label{subsec4}
Shifting gears,  we now discuss how to define the  \textit{proper distance}  in doubled-yet-gauged spacetime. First of all, we note that the usual infinitesimal one-form, $\rd x^{A}$, is neither doubled diffeomorphism covariant, $\delta(\rd x^{A})=\rd x^{B}\partial_{B}V^{A}\neq\rd x^{B}(\partial_{B}V^{A}-\partial^{A}V_{B})$,~\eqref{diff1}, \eqref{diff2},  nor  coordinate gauge symmetry invariant, $\rd\Delta^{A}=\rd x^{B}\partial_{B}\Delta^{A}\neq 0$, \eqref{CGS}.  Thus, the naive contraction with the DFT-metric, $\rd x^{A}\rd x^{B}\cH_{AB}$, cannot give  any sensible definition of  the  proper length.   To cure the problem,  one needs to  gauge $\rd x^{A}$  explicitly,   introducing  a connection, $\cA^{A}$,  which should satisfy the same property as the  coordinate gauge symmetry generator,  $\Delta^{A}$~\eqref{CGS},
\be
\ba{ll}
\rD x^{A}:=\rd x^{A}-\cA^{A}\,,\quad&\quad \cA^{A}\partial_{A}=0\,.
\ea
\label{gaugedDx}
\ee
Provided the connection  transforms appropriately,  $\rD x^{A}$ becomes a well-behaved  vector of  stringy geometry, meaning doubled  diffeomorphism covariant and  coordinate gauge symmetry invariant~\cite{Lee:2013hma}.  We propose then to define  the proper distance in doubled-yet-gauged spacetime  by a  \textit{path integral}~\cite{JHPBanff},
\be
\left|\left|\, x_{1}\,,\, x_{2}\,\right|\right|:=-\ln\left[\,\int \cD\cA\,\exp\left(-\int_{1}^{2}\sqrt{\rD x^{A}\rD x^{B}\cH_{AB}}\,\right)\,\right]\,.
\label{Length}
\ee
By  letting    $\tpartial^{\mu}\,{\equiv0}$ and $\cA^{A}\,{\equiv(A_{\mu},0)}$, we get   $\,\rD x^{A}\equiv(\rd\tx_{\mu}-A_{\mu},\rd x^{\nu})$. That is to say,   only the half of the doubled coordinates, \textit{i.e.} $\tx_{\mu}$ directions,  are gauged.   Furthermore, with the Riemannian parametrization of the DFT-metric, we get
\be
\rD x^{A}\rD x^{B}\cH_{AB}
\equiv\rd{x}^{\mu}\rd{x}^{\nu}g_{\mu\nu}+
\left(\rd{\tx}_{\mu}-A_{\mu}+\rd{x}^{\rho}B_{\rho\mu}\right)
\left(\rd{\tx}_{\nu}-A_{\nu}+\rd{x}^{\sigma}B_{\sigma\nu}\right)
g^{\mu\nu}\,.
\label{RH}
\ee
Thus, after integrating out the auxiliary  connection, $A_{\mu}$,  
our proposal~\eqref{Length}  reduces,  at least classically,   to the conventional   proper distance in Riemannian geometry,  ${\displaystyle{\int_{1}^{2}}}\sqrt{\rd{x}^{\mu}\rd{x}^{\nu}g_{\mu\nu}}\,$,   which is certainly independent of the    gauged  $\tx_{\mu}$ coordinates: 
\be
\left|\left|\, x^{A}_{1}\,,\, x^{A}_{2}\,\right|\right|=
\left|\left|\, x^{\mu}_{1}\,,\, x^{\mu}_{2}\,\right|\right|\,.
\ee
In this way,  indeed  the   formula~\eqref{Length} measures    the distance between two `gauge orbits'.

The exponent in \eqref{Length}   sets    immediately  the  action for a point particle  propagating in the doubled-yet-gauged spacetime, or its  square root free einbein formulation~\cite{Ko:2016dxa}.  As a consequence,  after the auxiliary connection being integrated out,  the point particle couples to $g_{\mu\nu}$ only. That is to say,  {both $\phi$ and $B_{\mu\nu}$ are \textit{dark} to particles: they are hard to detect by any  experiment   based on  particle concept.} Further, the point particle follows the geodesic  defined  \textit{not} in the Einstein frame \textit{but} in the string frame. Namely,  \textit{the Equivalence Principle is  restored  in the  {string frame}  for a point particle.}  This preferred choice of the frame is all due to the fundamental symmetries  of Stringy Gravity.

The above path integral definition of the proper length   can be easily generalized   to (Nambu-Goto type) area and volume, which in turn  gives the doubled-yet-gauged actions  for   string or  Green-Schwarz superstring~\cite{Lee:2013hma,Park:2016sbw}. In particular,  with auxiliary worldsheet metric, $h_{ij}$, and  the stringy Christoffel symbol~(\ref{Gammao}),   the Euler-Lagrange equations  of the string action  reduce, upon Riemannian backgrounds (\textit{c.f.~}non-Riemannian~\cite{Morand:2017fnv}), to 
\be
\textstyle{\frac{1}{\sqrt{-h}}}\partial_{i}\left(\sqrt{-h}\cH_{AB} D^{i}x^{B}\right)
+\Gamma_{ABC}\left(\brP^{B}{}_{D} D_{i}x^{D}\right)\left(P^{C}{}_{E}D^{i}x^{E}\right)=0\,,
\label{Sgeodesic}
\ee
which can be  viewed    as the stringy analogue    of the    point particle geodesic equation~(\ref{geodesic}).

\newpage

\section*{Acknowledgements} 
This note is based on series of works  by the author   in collaborations  with Imtak Jeon, Kanghoon Lee,  Soo-Jong Rey, Yuho Sakatani,  Yoonji Suh, Wonyoung Cho, Jose Fern\'andez-Melgarejo,
    Woohyun Rim, Sung Moon Ko, Charles Melby-Thompson,  Rene Mey\'er,  Minwoo Suh, Kang-Sin Choi,  Kevin Morand \,and\, Xavier Bekaert, as well as  discussions with Sang-Hyeon Ahn, David Berman, Robert Brandenberger, Martin Cederwall, Jaewon Lee,  Chan Park, Hyun Seok Yang and Sang Heon Yi. The author wishes to thank them all. This work was  supported by  the National Research Foundation of Korea   through  the Grant NRF-2016R1D1A1B01015196.


\end{document}